\documentclass{ws-ijmpd}

\begin{document}

\markboth{Armando Bernui, Marcelo J. Rebou\c{c}as}
{Non-Gaussianity of the HILC foreground-reduced three-year CMB map}

%
\catchline{}{}{}{}{}
%

\title{\uppercase{Non-Gaussianity in the HILC foreground-reduced three-year WMAP CMB map}}

\author{ARMANDO BERNUI\footnote{Permanent address: ICE, Universidade Federal de Itajub\'a,
MG, Brasil} \ \ AND \  MARCELO J. REBOU\c{C}AS}

\address{Centro Brasileiro de Pesquisas F\'{\i}sicas, Rua Dr. Xavier Sigaud 150\\
22290-180 Rio de Janeiro -- RJ,
Brasil\\
bernui@cbpf.br, reboucas@cbpf.br}

\maketitle

\begin{history}
\received{Day Month Year}
\revised{Day Month Year}
\comby{Managing Editor}
\end{history}

\begin{abstract}

A detection or nondetection of primordial non-Gaussianity in the CMB data
is essential not only to test alternative models of the physics of the early
universe but also to discriminate among classes of inflationary models.
Given this far reaching consequences of such a non-Gaussianity detection
for our understanding of the physics of the early universe, it is important
to employ alternative indicators in order to have further information about the
Gaussianity features of CMB that may be helpful for identifying their origins.
In this way, a considerable effort has recently gone into the design of
non-Gaussianity indicators, and in their application in the search for
deviation from Gaussianity in the CMB data.
Recently we have proposed two new large-angle non-Gaussianity indicators
which provide measures of the departure from Gaussianity on large angular
scales. We have used these indicators to carry out analyses of Gaussianity
of the single frequency bands and of the available foreground-reduced
\emph{five-year} maps with and without the \emph{KQ75} mask.
Here we extend and complement these studies by performing a new analysis of
deviation from Gaussianity of the \emph{three-year} harmonic ILC (HILC)
foreground-reduced  full-sky and \emph{KQ75} masked maps obtained from
WMAP data. We show that this full-sky foreground-reduced maps presents a
significant deviation from Gaussianity, which is brought down to a level
of consistency with Gaussianity when the \emph{KQ75} mask is employed.
\end{abstract}


\section{Introduction}

A detection or nondetection of primordial non-Gaussianity in the CMB data is
crucial not only to discriminate inflationary models but also to test
some alternative scenarios for the physics of the early universe.
However, the extraction of primordial non-Gaussianity is a difficult
enterprise since several effects of non-primordial nature can produce
non-Gaussianity in the CMB data.  
Clearly the study of detectable non-Gaussianities in the WMAP data must
take into account that they may have non-cosmological origins as, for
example, unsubtracted foreground contamination, unconsidered point sources
emission and systematic errors.\cite{Chiang-et-al2003,Naselsky-et-al2005,Cabella-et-al2009}
Deviation from Gaussianity may also have a cosmic topology origin
(see, e.g., the review articles Refs.~\refcite{CosmTopReviews} and  related
Refs.~\refcite{TopDetec}). 
If, on the one hand, different statistical tools can in principle provide
information about distinct forms of non-Gaussianity, on the other hand
one does not expect that a single statistical estimator can be sensitive
to all possible forms of non-Gaussianity in CMB data.
It is therefore important to test CMB data for Gaussianity by using
different statistical indicators to shed some light on its possible causes.
In view of this, a great deal of effort has recently gone into verifying the
existence of non-Gaussianity by employing several statistical
estimators.\cite{Some_non-Gauss-refs}

Recently have we proposed\cite{Bernui-Reboucas2009a} two new large-angle
non-Gaussianity indicators, based on skewness and kurtosis of large-angle
patches of CMB maps, which provide measures of the departure from Gaussianity
on large angular scales.
We used these indicators to search for the large-angle deviation from
Gaussianity in the three and five-year single frequency
K, Ka, Q, V, and W maps with and without a \emph{KQ75} mask.  We have
found strong deviation from Gaussianity in the unmasked maps, whereas
a \emph{KQ75} mask lowers significantly the level of non-Gaussianity
(for details see Ref.~\refcite{Bernui-Reboucas2009a}).

Motivated by the fact that most of Gaussianity analyses with Wilkinson
Microwave Anisotropy Probe (WMAP) data have been carried out by using CMB
frequency bands masked maps, and that sky cut can in principle induce bias
in Gaussianity analyses, in a more recent paper\cite{Bernui-Reboucas2009b}
we have carried out an analysis of Gaussianity of the available full-sky
foreground-reduced \emph{five-year} CMB
maps~\cite{ILC-5yr-Hishaw,HILC-Kim,NILC-Delabrouille} by
using the statistical indicators of Ref.~\refcite{Bernui-Reboucas2009a}.

We have shown that the available full-sky five-year foreground-reduced maps
present a significant deviation from Gaussianity, which varies with the
foreground-cleaning procedures. We have also shown that there is a substantial
reduction in the level of deviation from Gaussianity in these full sky maps
when a \emph{KQ75} mask is used.
Our main aim here is to extend and complement our previous work\cite{Bernui-Reboucas2009b}
by performing a similar analysis of Gaussianity of the \emph{three-year}
harmonic ILC (HILC) maps\cite{HILC-Kim} foreground-reduced  full-sky and
\emph{KQ75} masked maps. To this end, in the next section we give an account
of the large-angle non-Gaussianity indicators of Ref.~\refcite{Bernui-Reboucas2009a},
while in the last section we apply our indicators to perform a Gaussianity analysis
of the HILC three-year full-sky and \emph{KQ75} cut-sky maps, and present
our main results. Our principal conclusion is that the HILC
full-sky foreground-reduced maps presents a significant deviation from Gaussianity,
which is reduced to a level of consistency with Gaussianity when the \emph{KQ75}
mask is employed.

\section{Non-Gaussianity Indicators and Associated Maps}  \label{Indicators}

A constructive way of defining our non-Gaussianity indicators $S$
and $K$ (discrete functions defined on $S^2$) from CMB data
can be formalized through the following steps:%
\footnote{For a detailed discussion of the indicator briefly
presented here we refer the readers to
Ref.~\refcite{Bernui-Reboucas2009a} and Ref.~\refcite{Bernui-Reboucas2009b}.}
\begin{romanlist}
\item[{\bf i.}]
Take a finite set of points $\{j=1, \ldots ,N_{\rm c}\}$ homogeneously distributed
on the CMB  celestial sphere $S^2$ as the centers of spherical caps of a given
aperture $\gamma$; and calculate for each cap $j$ the skewness  and kurtosis 
given, respectively,  by
\begin{equation}
S_j   \equiv  \frac{1}{N_{\rm p} \,\sigma^3_{\!j} } \sum_{i=1}^{N_{\rm p}}
\left(\, T_i\, - \overline{T_j} \,\right)^3
\quad \mbox{and} \quad
K_j   \equiv  \frac{1}{N_{\rm p} \,\sigma^4_{\!j} } \sum_{i=1}^{N_{\rm p}}
\left(\,  T_i\, - \overline{T_j} \,\right)^4 - 3 \,,
\end{equation}
where $N_{\rm p}$ is the number of pixels in the $j^{\,\rm{th}}$ cap,
$T_i$ is the temperature at the $i^{\,\rm{th}}$ pixel, $\overline{T_j}$ is
the CMB mean temperature of the $j^{\,\rm{th}}$ cap, and $\sigma$ is the
standard deviation.
Clearly, the numbers $S_j$ and $K_j$ obtained in this way for each cap
can be viewed as a measure of non-Gaussianity in the direction of
the center of the cap $(\theta_j, \phi_j)$.
\item[{\bf iii.}]
Patching together the $S_j$ and $K_j$ values for each spherical cap,
one obtains our indicators, i.e., discrete functions $S = S(\theta,\phi)$
and $K = K(\theta,\phi)$ defined over the celestial sphere, which can be
used to measure the deviation from Gaussianity as a function of the angular
coordinates $(\theta,\phi)$. The Mollweide projection of skewness and kurtosis
functions $S = S(\theta,\phi)$ and $K = K(\theta,\phi)$  are nothing but
skewness and kurtosis maps, hereafter referred to them as $S-$map
and $K-$map, respectively.
\end{romanlist}

Clearly, the discrete functions $S = S(\theta,\phi)$ and $K = K(\theta,\phi)$
can be expanded into their spherical harmonics in order to determine their power
spectra $S_{\ell}$ and $K_{\ell}$.
Thus, for example, for the skewness one has
$S (\theta,\phi) = \sum_{\ell=0}^\infty \sum_{m=-\ell}^{\ell}
b_{\ell m} \,Y_{\ell m} (\theta,\phi)$ and
$S_{\ell} = (2\ell+1)^{-1}\sum_m |b_{\ell m}|^2$. 
Similar expressions obviously  hold for the kurtosis $K = K(\theta,\phi)$.

\section{Main Results and Conclusions}

In this section we shall report the results of our Gaussianity
analysis performed with   $S = S(\theta,\phi)$ and $K = K(\theta,\phi)$
indicators calculated from the foreground reduced HILC  full-sky and $KQ75$
masked maps computed from three-year WMAP data.%
\footnote{We note that in the analysis of Gaussianity with the \emph{KQ75} masked
maps the implementation of the mask is made by removing the pixels
inside the masked regions from the set of pixels of the each cap whose
intersection with the mask is not empty. Thus, the values $S_j$ and $K_j$ for
a  $j^{\,\rm{th}}$ cap (with pixels in the mask region) are calculated with
small number $N_{\rm p}$ of pixels.}

\begin{figure*}[htb!]
\begin{center}
\includegraphics[width=6cm,height=3.8cm]{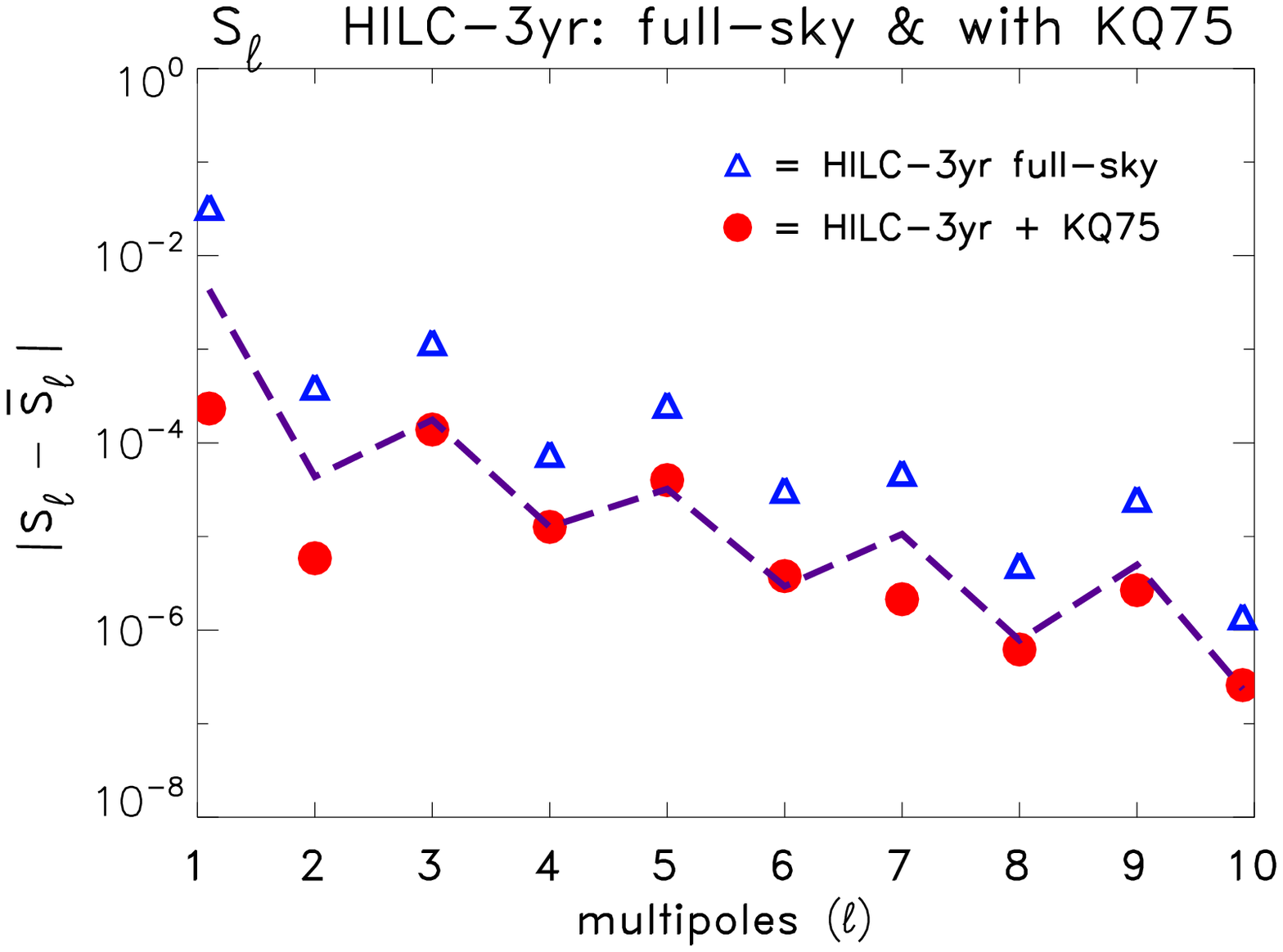}  
\hspace{0.3cm}
\includegraphics[width=6cm,height=3.8cm]{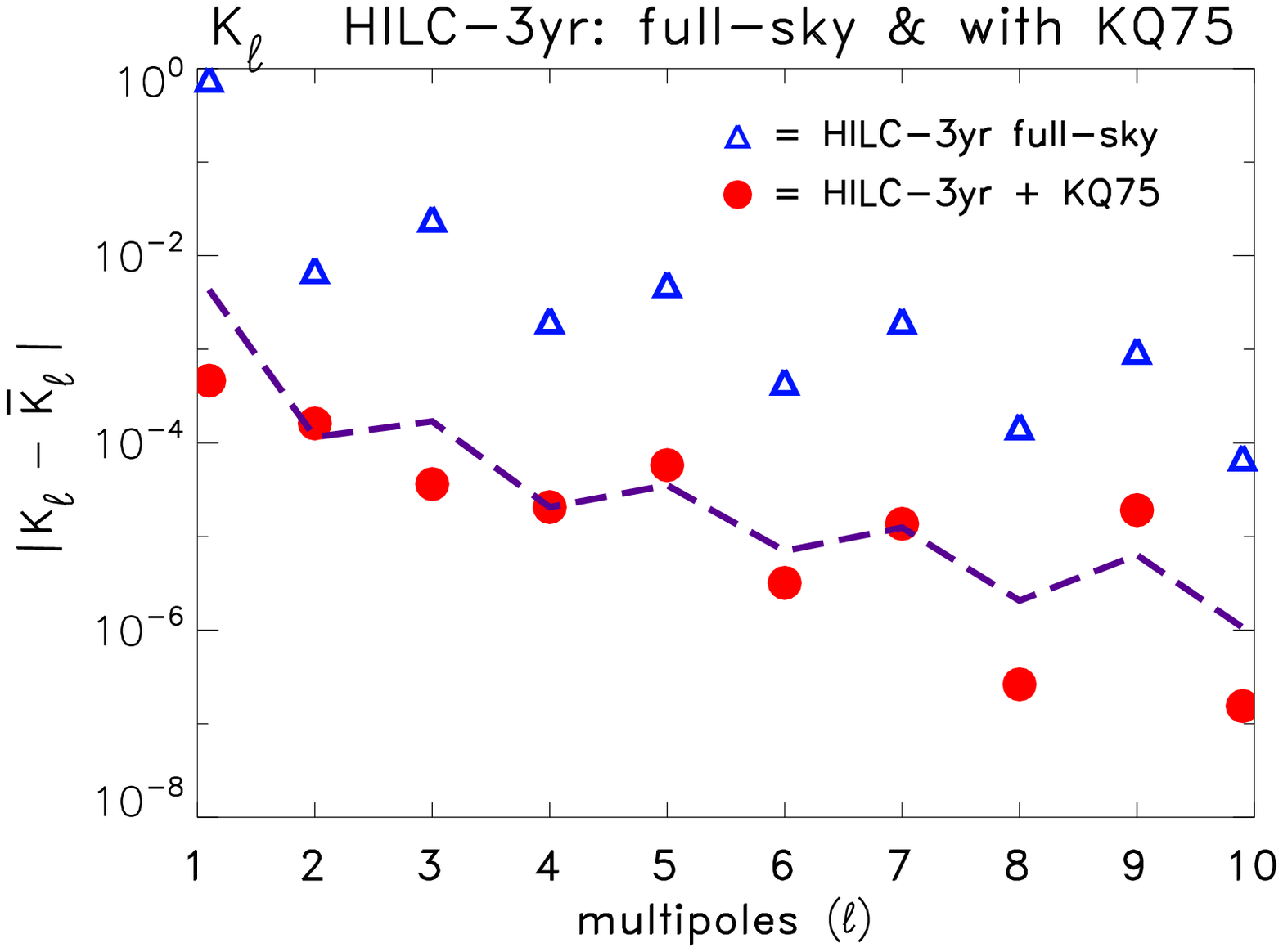}   
\caption{\label{Fig1}
Low $\ell $ \emph{differential} power spectra of skewness
$|S_{\ell} - \overline{S}_{\ell}|$ (left) and kurtosis (right)
$|K_{\ell} - \overline{K}_{\ell}|$
calculated from the  foreground-reduced HILC full-sky and
\emph{KQ75} masked maps.
The $95\%$ confidence level (obtained from Monte-Carlo Gaussian maps)
is indicated by the dashed line. \vspace{-0.6cm} }
\end{center}
\end{figure*}

To minimize the statistical noise, in the calculations
of $S-$map and $K-$map from the HILC  foreground-reduced three-year map,
we have scanned the celestial sphere with $12\,288$  spherical caps of
aperture  $\gamma = 90^{\circ}$, centered at points homogeneously
generated on the two-sphere by using  HEALPix\cite{Gorski-et-al-2005}.

Figure~\ref{Fig1} shows the differential power spectrum of the skewness
$S_{\ell}$ (left panel) and kurtosis $K_{\ell}$ (right panel) indicators
for $\,\ell=1,\,\,\cdots,10\,$, calculated from \emph{full-sky} and
\emph{KQ75} \emph{cut-sky} three-year foreground-reduced HILC maps.
The $95\%$ confidence level, obtained from $S$ and $K$ maps calculated
from Monte-Carlo (MC) statistically Gaussian CMB maps, is indicated
in this figure.%
\footnote{For details on the calculation of these (data and MC) maps and the
associated power spectra we refer the readers to Ref.~\refcite{Bernui-Reboucas2009a}
and Ref.~\refcite{Bernui-Reboucas2009b}.}
To the extent that the deviations $|S_{\ell} - \overline{S}_{\ell}|$ and
$|K_{\ell} - \overline{K}_{\ell}|$ for these maps are
not within $95\%$ of the mean MC value, Fig.~\ref{Fig1} shows an
important deviation from Gaussianity in full-sky foreground-reduced HILC
three-year map. This figure also shows a significant reduction in the level
of large-angle deviation from Gaussianity when the \emph{KQ75} mask is
used.

To have an overall assessment power spectra $S_\ell$ and $K_\ell$
calculated from the HILC foreground-reduced three-year full and cut map,
we have performed a $\chi^2$ test to find out the goodness of fit for
$S_{\ell}$ and $K_{\ell}$ multipole values as compared to the expected
multipole values obtained from $S$ and $K$ maps calculated
from Monte-Carlo (MC) statistically Gaussian CMB maps.
This gives a number for each case that quantifies collectively the
deviation from Gaussianity.
For the power spectra $S_\ell$ and  $K_\ell$ we found the values
given in Table~\ref{table1} for the ratio $\chi^2/\text{dof}\,$
(dof stands for degrees of freedom) for the power spectra calculated
from three-year HILC foreground-reduced full-sky and cut-sky maps.

\begin{table}[th]
\tbl{$\chi^2$ test goodness of fit for $S_{\ell}$ and $K_{\ell}$
calculated from the HILC  full-sky and cut-sky three-year maps as compared with
the expected values $\overline{S}_{\ell}$ and  $\overline{K}_{\ell}$ obtained
from MC maps  .}
{\begin{tabular}{@{}lcc@{}} \toprule 
Map  & $\chi^2$ for $S_{\ell}$  & $\chi^2$ for $K_{\ell}$  \\
\colrule
HILC full-sky           & $1.6 \times 10^3$  & $   8.8 \times 10^5 $   \\
HILC $KQ75$ cut-sky     & 0.8                & 1.5   \\
\botrule 
\end{tabular} \label{table1}}
\end{table}

Clearly, the greater is the values for $\chi^2/\text{dof}\,$  the smaller
are the $\chi^2$ probabilities, that is the probability that the power
spectra $S_{\ell}$ and $K_{\ell}$ and the expected  MC power spectra agree.
Thus, from  Table~\ref{table1} is one concludes that the HILC presents the
substantial level of deviation from Gaussianity as detected by the indicators,
which is reduced to a level that can be considered consistent with Gaussianity
when the \emph{KQ75} mask is employed.

Finally we note that the relative deviation of the full-sky power spectrum
from the cut-sky spectrum can be calculated with no reference to the Gaussian
MC spectra.
To this end, we have performed a $\chi^2$ test to find out the goodness
of fit for $S_{\ell}$ and $K_{\ell}$ multipole values for the full-sky
maps as compared to the corresponding cut-sky values. For this relative
assessment of power spectra $S_\ell$ and $K_\ell$ we have found
that $\chi^2/\text{dof}\,$ are $1.4 \times 10^3$ and $7.2 \times 10^5$.
These values make apparent the significant effect of the mask in the
reduction of the deviation from Gaussianity in the full-sky HILC
three-year map, and give information 
on reliability of the HILC full-sky foreground-reduced three-year
map as Gaussian reconstruction of the whole CMB sky.

\section*{Acknowledgments}
This work is supported by Conselho Nacional de Desenvolvimento
Cient\'{\i}fico e Tecnol\'{o}gico (CNPq) - Brasil, under grant No. 472436/2007-4.
M.J.R. and A.B. thank CNPq  for the grants under which this work was carried out.
We are also grateful to A.F.F. Teixeira for reading the manuscript
and indicating the omissions and misprints.
We acknowledge the use of the Legacy Archive for Microwave Background Data
Analysis (LAMBDA). Some of the results in this paper were derived using the HEALPix
package.\cite{Gorski-et-al-2005}

\end{document}